\documentclass[aps,twocolumn,superscriptaddress,showpacs]{revtex4}
\usepackage{graphicx}

\begin{document}

\title{
Unfolding the electronic structure of Ca$_{10}$(Fe$_{1-x}$Pt$_x$As)$_{10}$(Pt$_n$As$_8$) 
}
\author{Tom Berlijn}
\affiliation{Center for Nanophase Materials Sciences and Computer Science and Mathematics Division, Oak Ridge National Laboratory, Oak Ridge, TN 37831-6494, USA}
\date{\today}

\begin{abstract}
The iron platinum arsenides Ca$_{10}$(Fe$_{1-x}$Pt$_x$As)$_{10}$(Pt$_n$As$_8$) are the first Fe based superconductors with metallic spacer layers. 
Furthermore they display a large variation in their critical temperatures depending on the amount of Pt in their spacer layers: $(n=3,4)$.
To gain more insight into the role of the spacer layer the electronic structures of the iron platinum arsenides are represented in the momentum space of the underlying Fe sublattice using a first principles unfolding method.
We find that Ca$_{10}$(FeAs)$_{10}$(Pt$_4$As$_8$), contrary to Ca$_{10}$(FeAs)$_{10}$(Pt$_3$As$_8$), shows a net electron doping and a non-negligible interlayer coupling. 
Both effects could account for the difference in the critical temperatures. 
\end{abstract}


\maketitle

The common building blocks of the Fe based superconductors (FeSCs) are square lattices of Fe atoms with tetrahedrally surrounding As or Se atoms. 
What distinguishes the different families among them are the spacer layers. 
Some have none like FeSe. 
Others have spacer layers consisting of alkali atoms like LiFeAs, alkali-earth atoms like BaFe$_2$As$_2$, or rare earth oxides like LaFeAsO.
All these spacer layers are ionic and insulating. 
The newly discovered~\cite{skakiya,nni,cloehnert} family of iron platinum arsenides, Ca$_{10}$(Fe$_{1-x}$Pt$_x$As)$_{10}$(Pt$_3$As$_8$) (1038) and Ca$_{10}$(Fe$_{1-x}$Pt$_x$As)$_{10}$(Pt$_4$As$_8$) (1048), contains two types of spacer layers: a Ca and a PtAs layer. 
From the covalent nature of the PtAs layer ~\cite{skakiya} and from the Zintl concept of electron counting~\cite{nni,cloehnert} it was reasoned that the PtAs layer could be metallic.  
Indeed density functional theory calculations found the presence of Pt weight~\cite{cloehnert,irshein} and even Pt bands~\cite{mneupane,hnakamura} at the Fermi surface. 
Early Angle Resolved Photoemission Spectroscopy (ARPES) measurements~\cite{mneupane,sthirupathaiah} were not able to distinguish the presence of Pt bands.
However a  more recent ARPES experiment ~\cite{xpshen} on 1048 revealed several electron pockets around the extended Brillouin zones (BZs) of the PtAs layer.
That makes the iron platinum arsenides the first Fe based superconductors with metallic spacer layers.

An even more important aspect of the iron platinum arsenides is that they display a large difference in their optimal critical temperature ($T_c$) depending on the amount of Pt in the spacer layer.
While 1038 has a relatively low optimal $T_c$ between 10K and 15K ~\cite{skakiya,nni,zjxiang,kcho,mdwatson,sthirupathaiah, ttamegai}, $T_c$'s up to 38K ~\cite{skakiya,nni,cloehnert,qpding,ttamegai,kikeuchi,sthirupathaiah} have been reported for 1048. 
The fact that such a small change in the composition can make such a big difference in the $T_c$ raises the question as to what the essential tuning parameter of the superconductivity is. 
Nohara \textit{et al.} ~\cite{mnohora} proposed that the crucial difference lies in the As-Fe-As angle being closer to that of a perfect tetrahedron in 1048 than in 1038.
Cava \textit{et al.} ~\cite{nni} noted that the out-of-plane Pt-As bonding in 1048 is stronger than in 1038 and argued that the interlayer coupling is responsible for the $T_c$ enhancement.
Johrendt \textit{et al.} ~\cite{cloehnert} argued that the extra Pt in the PtAs layer injects electrons in the FeAs layer without the need for disorder inducing (Fe,Pt) substitutions.
In a subsequent study some of these same authors showed that 20\% of (La,Ca) substitutions in 1038 could ramp up its $T_c$ to 30K ~\cite{tstuerzer} which further supported their case that ``clean'' out of plane doping is the key to the high $T_c$. 
The La doping induced superconductivity was reproduced in other studies as well~\cite{nni2,jskim}. 
Borisenko \textit{et al.} ~\cite{sthirupathaiah} reasoned from their ARPES measurements that the critical difference between 1048 and 1038 lies in the number of band-edge singularities. 
Further experimental and theoretical works are needed to determine which of these mechanisms are truly responsible for the high $Tc$ in 1048. 

One difficulty for the theoretical investigations of the iron platinum arsenides is that the structure of the PtAs layers breaks the translational symmetry of the Fe lattice. 
Consequently the unit cells of 1038 and 1048 form $\sqrt{10}\times\sqrt{10}$ supercells with respect to the underlying Fe lattice. 
For electronic structure calculations in particular this means that the bands of the iron platinum arsenides will be folded in a BZ that is ten times as small,
 thereby significantly reducing the information they contain. 
To overcome this general problem of supercell calculations a first principles unfolding method has recently been developed ~\cite{unfolding} in which the band structures are generalized to spectral functions. 
The unfolding method not only restores the connectivity of the folded bands but also facilitates a direct comparison with the experimental ARPES observations.
Furthermore this procedure visualizes the coupling between of the unperturbed band structure and the symmetry breaker, the PtAs spacer layer in this case. 

\begin{figure}
\includegraphics[width=1\columnwidth,clip=true]{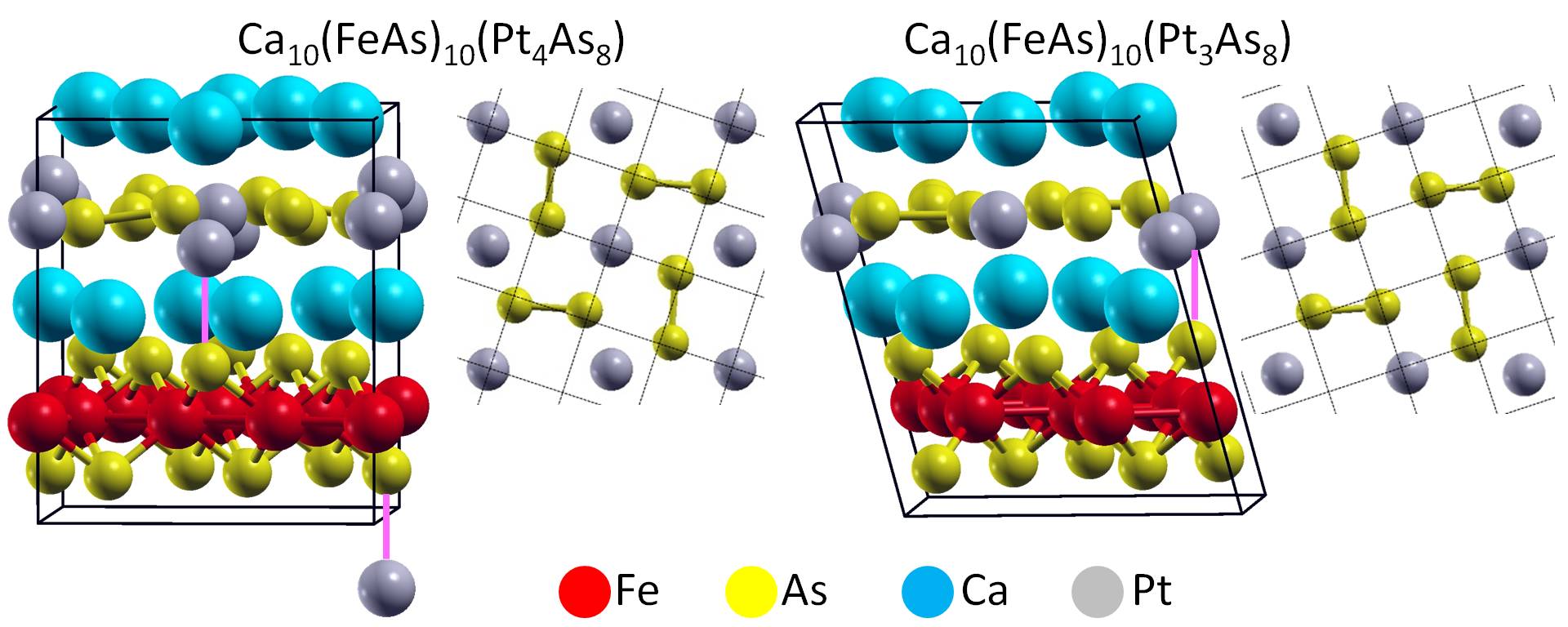}
\caption{\label{fig:fig1}
(color online) 
(Left/right) three dimensional view of the Ca$_{10}$(FeAs)$_{10}$(Pt$_4$As$_8$)/Ca$_{10}$(FeAs)$_{10}$(Pt$_3$As$_8$) unit cell together with top view of the PtAs layer in which the dotted lines indicate the Fe lattice.
The pink lines indicate the strong out-of-plane Pt-As bonds. Images were produced with the XCRYSDEN program ~\cite{kokalj}.
}\end{figure} 

\begin{figure*}[t]
\centering
\includegraphics[width=2\columnwidth,clip=true]{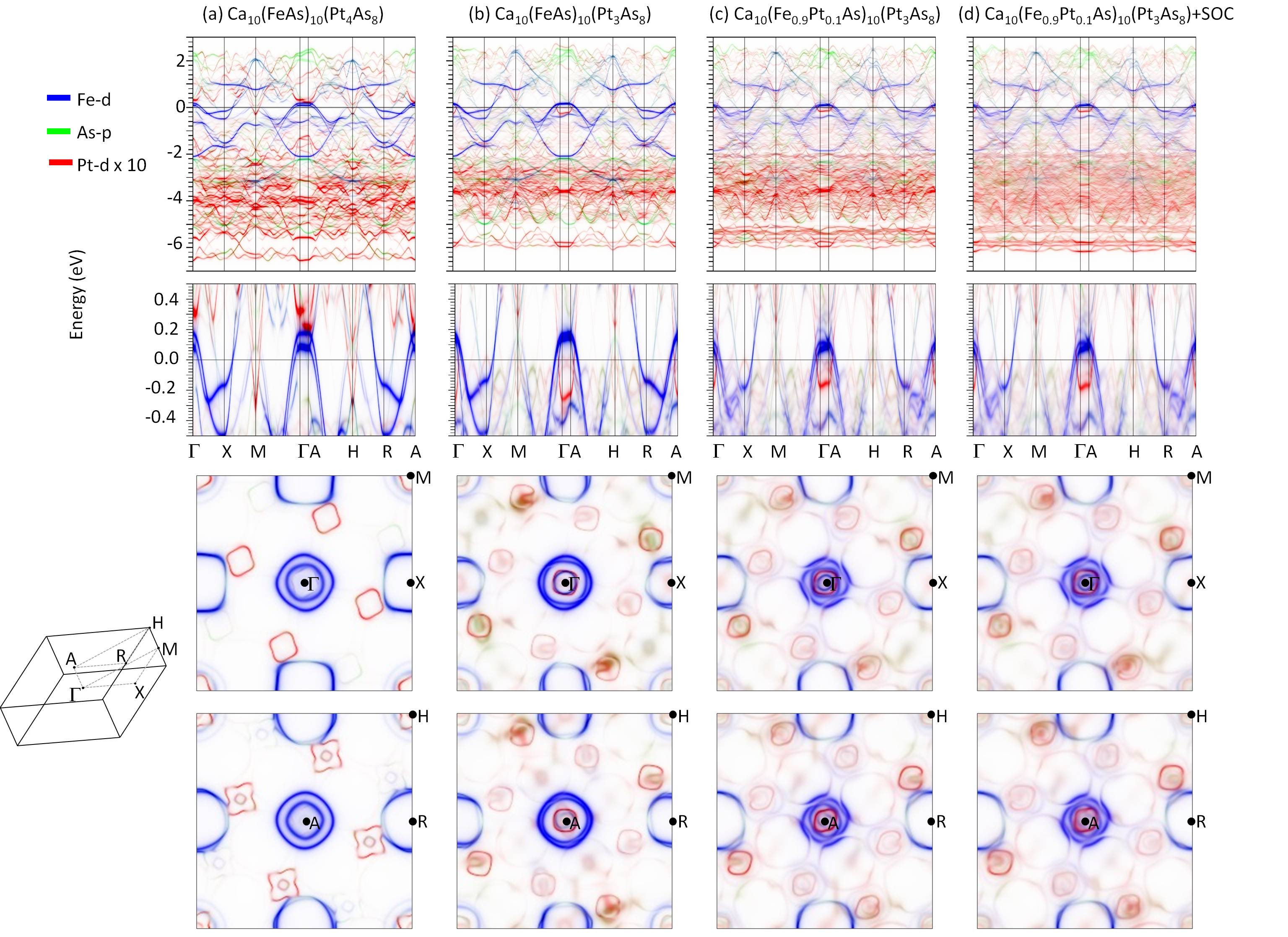}
\caption{
(color online)
Unfolded band structures and Fermi surfaces of (a) Ca$_{10}$(FeAs)$_{10}$(Pt$_4$As$_8$), (b) Ca$_{10}$(FeAs)$_{10}$(Pt$_3$As$_8$), and Ca$_{10}$(Fe$_{0.9}$Pt$_{0.1}$As)$_{10}$(Pt$_3$As$_8$) without (c) and with (d) treatment of the spin-orbit coupling in Pt in which the moment is constrained along the $a$-direction of the 1038 unitcell. The intensity of the Pt-$d$ bands has been enhanced by a factor of 10.
}
\label{fig:fig2}
\end{figure*}

In this paper the band structures of the iron platinum arsenides Ca$_{10}$(Fe$_{1-x}$Pt$_x$As)$_{10}$(Pt$_n$As$_8$) are unfolded to the Brillouin zone of the underlying Fe sublattice to better understand the role of the spacer layer for the superconductivity in these compounds.
We find that the Fe-$d$ bands are nominally undoped in Ca$_{10}$(FeAs)$_{10}$(Pt$_3$As$_8$) and electron doped by $\sim$0.1e/Fe in Ca$_{10}$(FeAs)$_{10}$(Pt$_4$As$_8$).   
Furthermore the Fe-$d$ electron pockets show an appreciable out-of-plane dispersion in Ca$_{10}$(FeAs)$_{10}$(Pt$_4$As$_8$) relative to Ca$_{10}$(FeAs)$_{10}$(Pt$_3$As$_8$) 
which reflects its stronger inter layer coupling in comparison.  
Both effects could account for their large difference in critical temperatures.
The (Fe,Pt) substitutions are found to induce strong disorder effects in the Fe-$d$ bands while the strong spin orbit coupling in the heavy Pt atoms displays no significant influence on the Fermi surface.   

The unfolded band structures are produced in three steps. 
First density functional theory calculations are performed.
To this end the WIEN2K implementation ~\cite{blaha} of the full potential linear augmented plane wave method was employed in the local density approximation.
The crystal structures are taken from Ref. ~\cite{nni}.
Next a Wannier transformation is performed using the projected Wannier function method~\cite{wannier}.
The low energy Hilbert space is taken within [-7,3]eV consisting of the Wannier orbitals of the Fe-$d$, Pt-$d$ and As-$p$ characters, 
resulting in a 119/124 orbital based tight binding Hamiltonian for 1038/1048.
Finally the eigenvectors and eigenvalues of the Hamiltonians are plugged into the unfolding formula of Ref. ~\cite{unfolding} to obtain the unfolded band structures and Fermi surfaces. 

Let us begin by reviewing the structures of 1038 and 1048 illustrated in Fig. \ref{fig:fig1}. 
Both unit cells consist of a stacking of a Ca$_5$ layer, a Pt$_n$As$_8$ (n=3,4) layer, another Ca$_5$ layer and a Fe$_{10}$As$_{10}$ layer. 
The main differences between 1038 and 1048 are as follows.
The Pt$_n$As$_8$ layer of 1038 is almost identical to that of 1048 accept for a Pt vacancy in the middle. 
As a consequence of the missing Pt the Ca layers in 1038 are stacked differently resulting a tilted structure of reduced triclinic symmetry. 
In addition the 1048 structure contains two out-of-plane As-Pt bonds (indicated by the pink lines) of length 3.08\AA  whereas 1038 contains only one of length 3.20\AA. 
This was pointed out by Cava \textit{et al.} ~\cite{nni} and used to argue that the interlayer coupling in 1048 is stronger than in 1038.
Finally let us consider the relation of the PtAs layer with respect to the Fe lattice.
In the top views in Fig. \ref{fig:fig1} the dotted lines indicate the projection of Fe sublattice onto the PtAs layer. 
From this we can see that the PtAs layer in 1038/1048 can be considered as an As lattice with two (As,Pt) substitutions and one/two Pt interstitial(s).
Obviously the Pt atoms strongly break the translation symmetry of the Fe lattice. 

The consequence of the strong breaking of the Fe lattice symmetry by the PtAs layer is clearly reflected in the unfolded band structures presented in Fig. \ref{fig:fig2}(a)(b), 
 in which the intensity of the Pt-$d$ bands has been enhanced by a factor of ten for better visibility.
If we focus our attention on the bands within the large energy window of [-7,3]eV we immediately notice the qualitative difference between the Fe-$d$ and Pt-$d$ bands.
Whereas the Fe-$d$ bands remain largely intact the Pt-$d$ bands are heavily reconstructed to the point that most of them appear more like an incoherent background without any momentum resolution. 
However if we zoom in on the bands within 0.5eV of the Fermi energy we note that the Pt-$d$ electron pockets, which were also observed in the folded band structures presented in Ref.~\cite{mneupane,hnakamura}, can still be resolved quit clearly. 
This is because their Fermi momenta are small enough such that these Pt-$d$ electron pockets cannot be nested by the reciprocal vectors of the 1048 and 1038 supercells.
Interestingly, we note in the Fermi surfaces of Fig. \ref{fig:fig2}(a) that the Pt-$d$ weight in 1048 vanishes in the electron pockets around the zone center and the four other even supercell reciprocal vectors. 
This however is not in contradiction with the observation of the electron pocket around the zone center in the recent ARPES experiment~\cite{xpshen}. 
The loss of Pt-$d$ weight in these pockets is compensated by the gain of Fe-$d$ weight that, although less visible in the color scale of Fig. \ref{fig:fig2}, is of similar intensity.
If we look at the bands of 1048 slightly above the Fermi energy we can see another electron pocket around 300meV which does have strong Pt-$d$ weight at the zone center.

Next let us investigate the degree of interlayer coupling by evaluating the out-of-plane dispersion. 
If we compare the Fermi surfaces at the $k_z=0$ and $k_z=\pi$ in Fig. \ref{fig:fig2}(a)(b) we see overall a relatively weak out-of-plane dispersion for both 1038 and 1048 in agreement with the experimental observations ~\cite{mneupane,sthirupathaiah,xpshen}.
However if we focus our attention on the Fe-$d$ electron pockets of 1048 in Fig. \ref{fig:fig2}(a) we can actually discern a non-negligible change in going from being square like at $k_z=0$ to circular like at $k_z=\pi$.
This is indicative of a stronger interaction between the FeAs and PtAs layers in 1048 compared to 1038. 
Another way to see that is to note the strong $k_z$ dependence of the Pt-$d$ electron pockets in 1048 which transform from square like into a flower like shape while dispersing along $k_z$.
The Pt-$d$ pockets in 1038 in Fig. \ref{fig:fig2}(b) on the other hand show very little out-of-plane dispersion.
So why is the interlayer coupling stronger in 1048 than in 1038?
Cava \textit{et al.} ~\cite{nni} proposed that the FeAs and PtAs layers are mainly connected via the strong PtAs bonds (indicated by the pink lines in Fig. \ref{fig:fig1}). 
In this picture the interlayer coupling in 1048 is stronger because it has two short PtAs bonds whereas 1038 has only one PtAs bond that is less short.
However from analyzing the Wannier function based tight binding Hamiltonians the bonds appear to be equally strong for both compounds (with a maximum hopping of $\sim$0.9eV between Pt-$d_{z^2}$ and As-$p_{z}$).   
Furthermore removing all the hopping elements in the PtAs bonds does not show any alteration of the Fermi surface of 1048. 
Most likely the interlayer coupling resides not in a select pair of bonds but in the collective of many.
For example the 1048 tight binding Hamiltonian shows 24 interlayer As bonds whose maximum hopping strengths range between 420 and 298meV.

Figure \ref{fig:fig2}(c)(d) show the effects of (Fe,Pt) substitution and spin-orbit coupling respectively. 
From comparing Fig. \ref{fig:fig2}(b)(c) we see that the (Fe,Pt) substitutions induce a large smearing of the Fe-$d$ bands in the [-2,2]eV range and also heavily reconstruct the hole pocket at the zone center. 
Evidently the (Fe,Pt) substitution affects the Fe-$d$ bands much more than the PtAs layer does. 
The microscopic reason for this is the strong impurity potential of the (Fe,Pt) substitution which from the tight binding Hamiltonian is found to contain a downward onsite energy shift of about $\sim$3eV.
Surprisingly the electron pockets at the Fermi level appear to be marginally influenced by this strong impurity potential.
To better understand this a more rigorous treatment of the disorder will be required~\cite{naxco2,mwhaverkort} that goes beyond the single impurity approximation used in this study. 
Qualitatively we can see that the (Fe,Pt) substitution dopes electrons into the system as the electron/hole pockets at the zone edge/center are reducing/increasing in size.
The amount of doped electrons however cannot be quantified from the Fermi surface as Luttinger's theorem~\cite{luttinger} breaks down in strongly disordered systems~\cite{tm122,kxfeyse2,mlfese,mwhaverkort}.
Another important effect to consider is the spin-orbit coupling in Pt given its large atom number. 
A strong effect of the spin-orbit coupling can be seen in the flat band around $\sim$-5.5eV which in splits by about $\sim$0.7eV as can be seen from comparing the top panels in Fig. \ref{fig:fig2}(c)(d).  
However the bottom panels in Fig. \ref{fig:fig2}(c)(d) show that the Pt-$d$ and Fe-$d$ bands at the Fermi energy are practically unaffected by the spin-orbit coupling.

\begin{figure}
\includegraphics[width=1\columnwidth,clip=true]{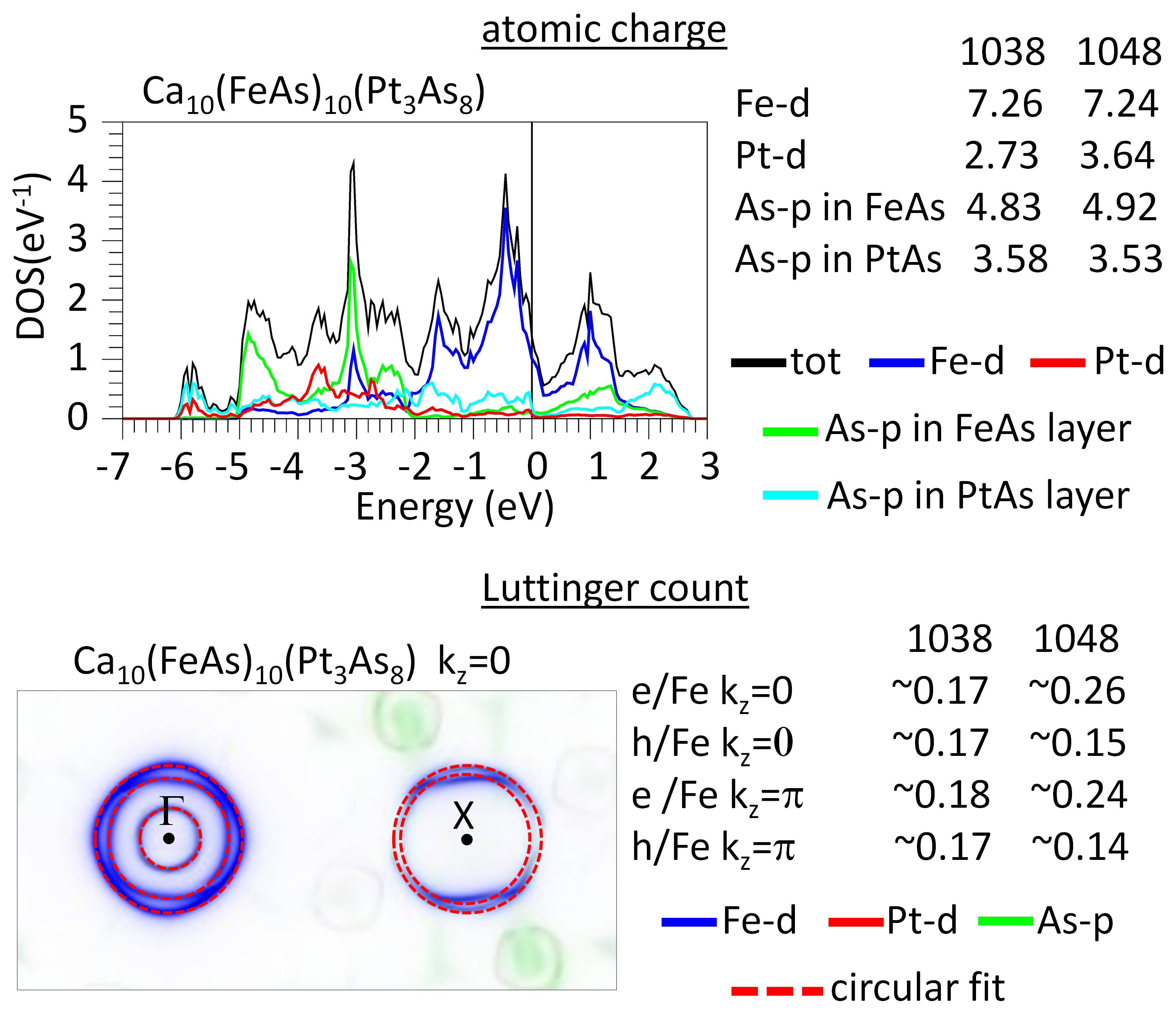}
\caption{\label{fig:fig3}
(color online) 
(Top right) atomic charges in Ca$_{10}$(FeAs)$_{10}$(Pt$_3$As$_8$)(1038) and Ca$_{10}$(FeAs)$_{10}$(Pt$_4$As$_8$) (1048) obtained from integration of the Wannier function resolved density of states.
(Top left) Wannier orbital resolved density of states of 1038.
(Bottom right) Luttinger count of carriers in the Fe electron and hole pockets obtained from circular fits.
(Bottom left) circular fits of electron and hole pockets of 1038. 
}\end{figure} 

Now let us go back to the iron platinum arsenides without (Fe,Pt) substitutions and investigate the doping effect of Pt in the PtAs spacer layer on the Fe-$d$ bands. 
In the top right of Fig. \ref{fig:fig3} the atomic charges per Fe are shown.
These are derived from integrating the Wannier function resolved density of states (such as shown in the top left of Fig. \ref{fig:fig3}) up to the Fermi level.
From comparing the atomic charges of 1048 with 1038 we see that roughly 9 out of 10 of the doped electrons remain in the Pt-$d$ shell, while the atomic charge in the Fe-$d$ orbitals remain almost constant.
This might give the impression that the additional Pt in 1048 does not induce any doping in the Fe-$d$ bands. 
However the Fe-$d$ bands hybridize strongly with these As-$p$ orbitals as can be seen from the large amount of Fe-$d$ weight within [-7,-2]eV and the large amount of As-$p$ weight in [-2,3]eV shown in the top left of Fig. \ref{fig:fig3}.
The amount of doped electrons that go into the Fe-$d$ bands are apparently matched by a decrease of Fe-$d$/As-$p$ hybridization such that the As-$p$ weight increases while the occupied Fe-$d$ weight remains constant.
From comparing the Fermi surfaces in Fig. \ref{fig:fig2}(a)(b) we can in fact clearly see that 1048 is electron doped compared with 1038 mainly from the increased size of the electron pockets.
To make an estimate of the number of doped electrons we perform a rough circular fit of the electron and hole pockets to quantify their enclosed volumes needed to perform a Luttinger count. 
From this we find first of all that in 1038 the number of holes is compensated by the number of electrons in agreement with the statement in ~\cite{tstuerzer} that 1038 is nominally undoped.
For 1048 we find that the net electron doping approximately equals $(0.26-0.15)\approx0.1$e/Fe  and (0.24-0.14)=0.1e/Fe in the $k_z=0$ and $k_z=\pi$ planes respectively. 
That number deviates significantly from 0.2e/Fe which is what one would have obtained from assuming the Pt$^{2+}$ valence state.
Nonetheless this estimate shows that the additional Pt in 1048 compared to 1038 induces a large electron doping in the Fe-$d$ bands.

In this study we noted that while both the (Fe,Pt) substitutions and the Pt$_4$As$_8$ layer electron dope the iron platinum arsenides, only the (Fe,Pt) substitutions induce a strong disorder effect on the Fermi surface. 
This finding supports the picture of Johrendt \textit{et al.} ~\cite{cloehnert,tstuerzer} that the role of the additional Pt in 1048 is to inject electrons without the need for the disorder inducing (Fe,Pt) substitutions. 
However there have also been reports~\cite{kikeuchi,skakiya} of $T_c$'s as high as 38K in 1048 in presence of large amounts of (Fe,Pt) substitutions of approximately 20\%. 
These studies seem to disagree with the idea that the disorder induced by the (Fe,Pt) substitutions is detrimental to the superconductivity. 
Alternatively the interlayer coupling could play an important role for the formation of both the superconducting order, as well as potentially competing magnetic and structural orders ~\cite{tzhou,nni2,tstuerzer2,pwgao} such as have recently been reported for 1038. 
On the one the hand the reduced dimensionality of a weakened interlayer coupling improves the nesting conditions of the Fermi surface, possibly leading to larger instabilities toward magnetic and structural orders and enhanced superconducting pairing due to larger spin fluctuations. 
On the other hand a small but finite degree of interlayer coupling can be instrumental since long range order cannot exists in pure two-dimensional systems, be it of magnetic~\cite{merminwagner} or superconducting~\cite{hohenberg} order.
High pressure experiments ~\cite{mnohora,pwgao} have shown that the superconductivity in 1038/1048 is induced/suppressed under the influence of pressure.
The authors of Ref. ~\cite{pwgao} argue that the pressure induced increased bandwidth of the PtAs layer causes a charge transfer to the FeAs layer. 
Yet, undoubtedly that same pressure will enhance the interlayer coupling as well. 
The findings in the literature as well as those in the present paper indicate that both doping and the interlayer coupling could play an important role for the superconductivity in the iron platinum arsenides. 
More theoretical and experimental studies will be desirable to study their relative importance. 
 
To summarize, in this paper the band structures and Fermi surfaces of the iron platinum arsenides have been unfolded to the Brillouin zone of the Fe lattice by means of first-principles calculations.
From applying Luttinger's theorem to the unfolded Fe-$d$ bands we find that Ca$_{10}$(FeAs)$_{10}$(Pt$_3$As$_8$) is a compensated semimetal while Ca$_{10}$(FeAs)$_{10}$(Pt$_4$As$_8$) displays a net electron doping of approximately 0.1e/Fe.
Furthermore the Fe-$d$ and Pt-$d$ electron pockets in Ca$_{10}$(FeAs)$_{10}$(Pt$_4$As$_8$) show a sizable out-of-plane dispersion whereas those in Ca$_{10}$(FeAs)$_{10}$(Pt$_3$As$_8$) do not.
This indicates that Ca$_{10}$(FeAs)$_{10}$(Pt$_4$As$_8$) has a stronger interlayer coupling than Ca$_{10}$(FeAs)$_{10}$(Pt$_3$As$_8$).
Both the electron doping and the enhanced interlayer coupling could explain why the optimal $T_c$ of Ca$_{10}$(Fe$_{1-x}$Pt$_x$As)$_{10}$(Pt$_4$As$_8$) is more than twice as high in comparison. 
The (Pt,Fe) substitutions are found to strongly reconstruct the hole pockets of Ca$_{10}$(FeAs)$_{10}$(Pt$_3$As$_8$) while the spin-orbit coupling in Pt shows no effect on the Fermi surface of this system. 

TB was supported as a Wigner Fellow at the Oak Ridge National Laboratory.
We would like to thank D. Inosov for suggesting this problem and W. Ku for use of his Wannier function code.

\end{document}